\documentclass[prb]{revtex4}

\newlength{\figwidth}
\usepackage{graphicx}

\newcommand{\cth}{\ensuremath{\left\langle \cos \theta \right\rangle}}
\newcommand{\csth}{\ensuremath{\left\langle \cos^2 \theta \right\rangle}}
\newcommand{\cible}{\ensuremath{\psi_{\mathrm{target}}}}

\begin{document}

\title{Optimal molecular alignment and orientation through rotational
  ladder climbing}

\author{Julien Salomon}
\email{salomon@ann.jussieu.fr}
\affiliation{Laboratoire Jacques-Louis Lions,
Universit\'{e} Pierre \& Marie Curie,
Bo\^{\i}te courrier 187, 75252 Paris Cedex 05,
France}

\author{Claude M. Dion}
\email{claude.dion@tp.umu.se}
\affiliation{Department of Physics, Ume{\aa} University, SE-901\,87 Ume{\aa}, Sweden}

\author{Gabriel Turinici}
\email{Gabriel.Turinici@dauphine.fr}
\affiliation{INRIA Rocquencourt, B.P. 105, 78153 Le Chesnay Cedex}
\affiliation{CERMICS-ENPC, Champs-sur-Marne, 77455 Marne-la-Vall\'{e}e Cedex,
France}

\begin{abstract}
  We study the control by electromagnetic fields of molecular
  alignment and orientation, in a linear, rigid rotor model.  With the
  help of a monotonically convergent algorithm, we find that the
  optimal field is in the microwave part of the spectrum and acts by
  resonantly exciting the rotation of the molecule progressively from
  the ground state, i.e., by rotational ladder climbing. This
  mechanism is present not only when maximizing orientation or
  alignment, but also when using prescribed target states that
  simultaneously optimize the efficiency of orientation/alignment and
  its duration.  The extension of the optimization method to
    consider a finite rotational temperature is also presented.
\end{abstract}

\date{\today}

\maketitle

\section{Introduction}

External fields can be used to manipulate molecules to achieve
molecular axis alignment or orientation. Here, alignment refers to
setting the molecular axis parallel to a laboratory fixed frame, while
orientation implies that the molecular axis has in addition the same
direction as the laboratory fixed frame.  These two goals have a wide
range of applications in fields such as chemical
reactivity,\cite{orient:brooks76} surface
processing,\cite{orient:tenner91,focus:mcclelland93} nanoscale
design,\cite{focus:seideman97b,manip:dey00} attosecond pulse
production,\cite{align:bandrauk03,align:denalda04} and quantum
information processing.\cite{align:lee04b}

Efficient alignment~\cite{align:sakai99,align:larsen99} and
orientation~\cite{orient:kanai01,orient:guerin02} can be achieved by
laser-induced adiabatic passage from field-free rotational states to
aligned pendular
states,\cite{pendul:zon75,align:friedrich95a,align:ortigoso99,%
  align:andryushin99,floquet:keller00a} but it is lost at the end of
the laser pulse.  Field-free alignment is possible by sudden
excitation using pulses much shorter than the rotational period of the
molecule.\cite{align:seideman95,align:seideman99,orient:henriksen99,%
  align:dion99,align:renard03} Achieving orientation with short pulses
is more difficult since spatial symmetry breaking is required to give
the direction of orientation, but it can be done using half-cycle
pulses\cite{hcp:dion01,orient:machholm01} or specially tailored laser
pulses.\cite{orient:dion02,orient:atabek03b} This has led to different
proposals for alignment and orientation using series of short impulses
(kicks).\cite{control:averbukh01,align:leibscher03,%
  orient:matos-abiague03,align:ortigoso04,align:leibscher04,align:lee04,%
    align:bisgaard04,align:sugny04a} See also the recent review of the
subject by Stapelfeldt and Seideman, and references
therein.\cite{align:stapelfeldt03}

The purpose of the present study is to find the electromagnetic fields
that produce the best possible orientation or alignment.  We start by
presenting in Sec.~\ref{sec:model} the rigid rotor model used to
describe the rotation of a linear molecule, along with the cost
functionals that describes the required control objectives, in terms
of both observables measuring orientation or alignment and target
states that embody the efficiency of orientation/alignment along with
its persistence.  

The optimization procedure itself is based on monotonically convergent
algorithms~\cite{control:tannor92,control:zhu98a,control:maday03a}
that are guaranteed to improve at each step the cost functional
chosen. The corresponding algorithm for the control of
alignment/orientation is presented in Sec.~\ref{sec:mca}.

As we will see in Sec.~\ref{sec:results}, the fields leading to
optimal orientation and alignment are in the microwave part of the
spectrum and lead to \emph{rotational ladder climbing}, i.e., the
molecule is resonantly excited successively from one rotational level
to the next.  The possibility of controlling rotational excitation by
ladder climbing using microwave fields was first proposed by Judson
\emph{et al.}~\cite{rot:judson90,control:judson92} A process similar
but resting on Raman excitation of ro-vibrational states with chirped
pulses has been used to create an optical centrifuge for
molecules.\cite{manip:karczmarek99,manip:villeneuve00,%
  rot:spanner01a,rot:spanner01b,rot:vitanov04}

\section{Model}
\label{sec:model}

\subsection{Time-dependent Schr\"{o}dinger equation}
\label{sec:tdse}

The dynamics of the molecule interacting with the electromagnetic
field is obtained by solving the time-dependent Schr\"{o}dinger equation
(TDSE).  We restrict ourselves to the case of a linear molecule in a
rigid rotor approximation, yielding the Hamiltonian (in atomic units,
$\hbar = 1$)
\begin{equation}
\hat{H} = B \hat{J}^2 - \mu_{0} \mathcal{E}(t) \cos\theta -
 \left[\left(\alpha_{\parallel}-\alpha_{\perp}\right) \cos^{2}\theta
  + \alpha_{\perp} \right]\frac{\mathcal{E}^{2}(t)}{2},
\label{eq:tdse}
\end{equation}
where $B$ is the rotational constant, $\hat{J}$ is the angular
momentum operator, $\theta$ is the polar angle positioning the
molecular axis with respect to the polarization vector of the linearly
polarized electric field of amplitude $\mathcal{E}(t)$, $\mu_{0}$ is
the permanent dipole moment, and $\alpha_{\parallel}$ and
$\alpha_{\perp}$ are the dipole polarizability components parallel and
perpendicular to the molecular axis, respectively.  Because of the
cylindrical symmetry about the field polarization axis, the motion
associated with the azimuthal angle can be separated and $M$, the
projection of the total angular momentum $J$ on the axis, is a good
quantum number ($\Delta M = 0$).  The TDSE~(\ref{eq:tdse}) is
solved numerically starting from the ground rotational (isotropic)
state $J=M=0$, using a basis set expansion of the wave function $\psi$
in terms of the spherical harmonics $Y_{J,M}$,
\begin{equation}
\psi(\theta,t) = \sum_{J=0}^{\infty} c_J(t) Y_{J,0}(\theta),
\label{eq:expand}
\end{equation} 
the $c_J$ being complex coefficients and the coupling terms due to
$\mu_{0}$ and $\alpha$ being then
analytical.\cite{orient:benhajyedder02}  For computational purposes,
only the first 10 terms in the sum in Eq.~(\ref{eq:expand}) are kept,
and we have checked that the results are not affected by using a
bigger basis.

Because of the presence of both the dipole moment $\mu$ and the
polarizability anisotropy $\Delta \alpha \equiv
\alpha_{\parallel}-\alpha_{\perp}$, the results are not
molecule-independent.  However, the role of the polarizability is
negligible for the fields obtained, making them applicable to any
linear molecule with a proper scaling of the amplitude and frequency
of the electric field.  The results will thus be presented with time
expressed in units of the rotational period $T_\mathrm{rot} = h/2B$,
the electrical field as $\mu_0 \mathcal{E}/B$, and energy as $E/B$.
The parameters actually used in the calculations are those for the HCN
molecule: $B = 6.6376 \times 10^{-6}$, $\mu = 1.1413$,
$\alpha_\parallel = 20.055$, and $\alpha_\perp = 8.638$ (all in atomic
units).

\subsection{Cost functional}

As we are seeking to optimize molecular orientation or alignment, our
cost functionals are based on their respective measure, the
expectation values $\cth$ and $\csth$.  A molecule will be oriented
when $\left| \cth \right| \sim 1$, with the sign indicating in which
direction it is pointing; an angular distribution symmetric with
respect to $\theta = \pi/2$ will yield a value of zero.  The
expectation value of $\cos^2 \theta$ is 1 when the molecule is
aligned, starting from $1/3$ for the isotropic case.

The first case we consider is a cost functional of the form
\begin{equation}
  \mathcal{J}_1(\mathcal{E}) = \left\langle \psi(t_\mathrm{f}) \right|
  \hat{O} \left| \psi(t_\mathrm{f}) \right\rangle -
  \int_{0}^{t_\mathrm{f}} \lambda(t) \mathcal{E}^2(t) \, dt,
\label{eq:costop}
\end{equation}
with $\hat{O}$ an operator chosen to be $\cos\theta + \hat{I}$ for
orientation and $\cos^2\theta + \hat{I}$ for alignment, the identity
operator $\hat{I}$ being used for convenience (e.g., it ensures that
$\hat{O}$ is positive) without modifying the extrema of
$\mathcal{J}_1$, and $t_\mathrm{f}$ the time at which the interaction
with the field ends.  The last term in Eq.~(\ref{eq:costop}) is a
penalization on the amplitude of the field, with
\begin{equation}
  \lambda(t) = 10^5 \left( \frac{t-t_\mathrm{f}/2}{t_\mathrm{f}/2}
  \right)^6+10^4. 
\label{eq:lambda}
\end{equation}
This imposes a strong constraint on the maximum amplitude of the
electric field, such that the field strengths are comparable to those
that can be achieved for half-cycle
pulses,\cite{hcp:you93,hcp:bucksbaum00} allowing a comparison with
previously published results for alignment and orientation using
kicks.\cite{hcp:dion01,orient:machholm01,control:averbukh01,%
  align:leibscher03,orient:matos-abiague03,align:ortigoso04,%
  align:leibscher04,align:lee04,align:bisgaard04,align:sugny04a} In
addition, the form of Eq.~(\ref{eq:lambda}) ensures a smoother, and
thus more realistic, turn-on and turn-off of the field.  We point out
that, in any case, this penalty term is not an essential ingredient of
the monotonic algorithm and can be relaxed (e.g., to allow fluences
achievable with laser pulses) or completely eliminated.

The downside of such a cost functional is that it takes into account
only the efficiency of the orientation/alignment, not its persistence.
Once the field is turned off, the free rotation of the molecule will
lead to the disappearance of the orientation/alignment as the
different $J$ components in the wave function dephase, followed by
revivals at intervals of one rotational period.\cite{align:seideman99}
The best orientation/alignment is obtained by confining the molecule
to a narrow angular distribution $\Delta \theta$, which corresponds to
exciting a broad rotational band $\Delta J$ by referring to an
uncertainty principle $\Delta J\cdot \Delta \theta \sim
\hbar$.\cite{align:stapelfeldt03} The problem is then that,
conversely, a broad rotational spectrum exhibits narrow features in
the time domain, i.e., the greater the orientation/alignment, the
shorter its duration.  A compromise has thus to be made, as can be
achieved by considering the best orientation/alignment possible for a
restricted maximum rotational excitation.  The procedure on how states
with such characteristics can be obtained is given in detail in
Refs.~\onlinecite{align:sugny04a,control:sugny05a}, where it can also
be seen that $J_{\mathrm{max}} = 4$ offers a good compromise, leading
to an orientation of $\cth \approx 0.906$ or an alignment of $\csth
\approx 0.837$, both lasting of the order of 1/10th of the rotational
period. The cost functional is now
\begin{equation}
  \mathcal{J}_2(\mathcal{E}) = 2\Re \left\langle \cible \left|
      \psi(t_\mathrm{f}) \right\rangle \right. -
  \int_{0}^{t_\mathrm{f}} \lambda(t) \mathcal{E}^2(t) \, dt,
\label{eq:costop_target}
\end{equation}
where $\cible$ denotes the target state corresponding to orientation
or alignment, as given in Table~\ref{tab:target}, and $\Re$ the real
part. Note that because of the norm conservation properties of the
Schr\"{o}dinger equation, the cost functional~(\ref{eq:costop_target}) has
the same minima and critical points as
\begin{equation}
  \mathcal{J}(\mathcal{E}) = - \left\| \cible - \psi(t_\mathrm{f})
  \right\|^2 - \int_{0}^{t_\mathrm{f}} \lambda(t) \mathcal{E}^2(t) \,
  dt, 
\end{equation}
which measures the distance between $\cible$ and $\psi(t_\mathrm{f})$.

In all cases, the time at which the field is turned off and the cost
functional measured is chosen as $t_\mathrm{f} = 9.5\times 10^6\
\mathrm{a.u.} \approx 20 T_\mathrm{rot}$ for the results presented
here.  Shorter durations lead to results either similar or less
significant.

\subsection{Monotonically convergent algorithm}
\label{sec:mca}

The algorithm used to find the optimal field is based on a general
class of monotonically convergent algorithms recently
proposed.\cite{control:maday03a} We present here the algorithm
associated to $\mathcal{J}_1$ and refer the reader to
Refs.~\onlinecite{control:maday03b,control:maday05} for a detailed
discussion the algorithm in a time-discretized framework.  At the
maximum of the cost functional $\mathcal{J}_1$, the Euler-Lagrange
critical point equations are satisfied; a standard way to write these
equations is to use a Lagrange multiplier $\chi(\theta,t)$ called
\emph{adjoint state}. The following critical point equations are thus
obtained:
\begin{equation}\label{critic}
\begin{array}{rclrcl}
i\partial_t\psi & = & \hat{H} \psi,& \quad \psi(0) & = & \psi_0, \\
i\partial_t\chi & = & \hat{H} \chi,& \quad \chi(t_\mathrm{f}) & = &
\hat{O}\biglb(\psi(t_\mathrm{f})\bigrb), \\
\multicolumn{6}{l}{
\lambda(t)\mathcal{E}(t) = -\Im \left\langle \chi(t) \right| \mu_0 \cos\theta
      + 2\mathcal{E}(t) \left(\Delta\alpha\cos^2\theta+\alpha_{\perp}
      \right) \left|\psi(t) \right\rangle,}
\end{array}
\end{equation}
where $\Im$ is the imaginary part of a complex number and $\psi_0$ the
initial state of the controlled system.

Given two fields $\mathcal{E}$ and $\mathcal{E}'$ and the
corresponding states $\psi$, $\psi'$ and adjoint states $\chi$,
$\chi'$ defined by Eq.~(\ref{critic}), one can write
\begin{eqnarray}
\Delta\mathcal{J}_1 & = &
\mathcal{J}_1(\mathcal{E}')-\mathcal{J}_1(\mathcal{E}) \nonumber \\
 & = & \left\langle \psi'(t_\mathrm{f})-\psi(t_\mathrm{f}) \right| \hat{O} \left|
   \psi'(t_\mathrm{f})-\psi(t_\mathrm{f}) \right\rangle \nonumber \\
 && + \int_0^{t_\mathrm{f}} \left[ \mathcal{E}'(t)-\mathcal{E}(t)
 \right] \Biggl\{ 2\Im \left\langle \psi'(t) \right| \mu_0\cos\theta
 \left| \chi(t) \right\rangle \nonumber \\
&& \phantom{+ \int_0^{t_\mathrm{f}}(} + \left[
  \mathcal{E}'(t)+\mathcal{E}(t) \right] \left[ 2\Im \left\langle
    \psi'(t) \right| \frac{\Delta\alpha\cos^2\theta+\alpha_{\perp}}{2}
  \left| \chi(t) \right\rangle - \lambda(t)\right] \Biggr\} \, dt.
\label{variation}
\end{eqnarray}
The first term of this sum is positive since both choices
$\hat{O}=\cos+ \hat{I}$ or $\hat{O}=\cos^2+ \hat{I}$ are
positive. Given $\mathcal{E}$, the integrand provides thus an implicit
criterion in terms of $\mathcal{E}'$, the satisfaction of which
guarantees the positivity of $\Delta \mathcal{J}_1$. An explicit
choice of $\mathcal{E}'$ can be exhibited: the integrand of
Eq.~(\ref{variation}) is a second-order polynomial with respect to
$\mathcal{E}'$ and for a large enough value of $\lambda(t)$ the
coefficient $2\Im \left\langle \psi'(t) \right|
\frac{\Delta\alpha\cos^2\theta+\alpha_{\perp}}{2} \left| \chi(t)
\right\rangle - \lambda(t)$ of $\mathcal{E}'^2(t)$ is negative. It has
thus a unique maximum, given by the cancellation of the
derivative. The value obtained by this method is
\begin{equation}
\mathcal{E}'(t) =
-\frac{\Im \left\langle \psi'(t) \right| \mu_0\cos\theta \left|
    \chi(t) \right\rangle}{2\Im \left\langle \psi'(t) \right|
  \frac{\Delta\alpha\cos^2\theta+\alpha_{\perp}}{2} \left| \chi(t)
  \right\rangle -\lambda(t)}.  
\end{equation}
The algorithm derived from the previous computations is then given by
the following procedure: given at step $k$ a field $\mathcal{E}^k$ and
its associated state $\psi^k$ and adjoint state $\chi^k$, compute
simultaneously $\mathcal{E}^{k+1}$, $\psi^{k+1}$ by
\begin{equation}
\left\{
\begin{array}{rcl}
  \mathcal{E}^{k+1}(t) & = & -\frac{\Im \left\langle \psi^{k+1}(t) \right|
    \mu_0\cos\theta \left| \chi^k(t) \right\rangle}{2\Im \left\langle
      \psi^{k+1}(t) \right| \frac{\Delta\alpha\cos^2\theta
      +\alpha_{\perp}}{2} \left| \chi^k(t) \right\rangle -\lambda(t)},
  \\
  i\partial_t \psi^{k+1}(t) & = & \left[
  B-\mu_0\mathcal{E}^{k+1}(t)\cos\theta
  -\frac{[\mathcal{E}^{k+1}(t)]^2}{2}
  \left(\Delta\alpha\cos^2\theta+\alpha_{\perp} \right) \right]
  \psi^{k+1}(t), 
  \\
  \psi^{k+1}(\theta,t=0) & = & \psi_0(\theta).
\end{array}\right.
\label{eq:algo}
\end{equation}
Then compute backward evolution of $\chi^{k+1}$ by 
\begin{equation}
\left\{
  \begin{array}{rcl}
    i\partial_t \chi^{k+1}(t) & = & \left[
      B-\mu_0\mathcal{E}^{k+1}(t)\cos\theta
      -\frac{[\mathcal{E}^{k+1}(t)]^2}{2} \left(
        \Delta\alpha\cos^2\theta+\alpha_{\perp} \right) \right]
    \chi^{k+1}(t), \\  
    \chi^{k+1}(\theta,t_\mathrm{f}) & =& \hat{O} \biglb(
    \psi^{k+1}(\theta,t_\mathrm{f}) \bigrb).
  \end{array}\right.
\end{equation}
The arguments above show that 
\begin{equation}
\mathcal{J}_1(\mathcal{E}^{k+1})\geq \mathcal{J}_1(\mathcal{E}^k).
\end{equation}

\section{Results}
\label{sec:results}

\subsection{Optimizing orientation}

\begin{figure}
\includegraphics[width=\figwidth]{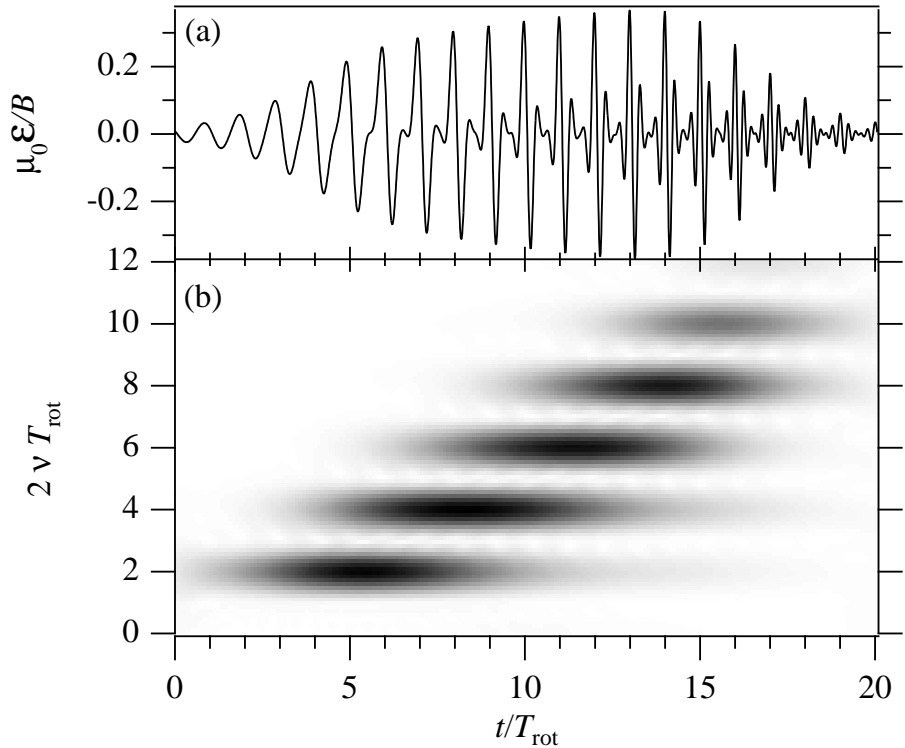}
\caption{(a) Electric field obtained with criterion $\mathcal{J}_1$
  for the optimization of $\cth$.  (b) Short-time Fourier transform of
  the field in (a).}
\label{fig:champ_cos}
\end{figure}

\begin{figure}
\includegraphics[width=\figwidth]{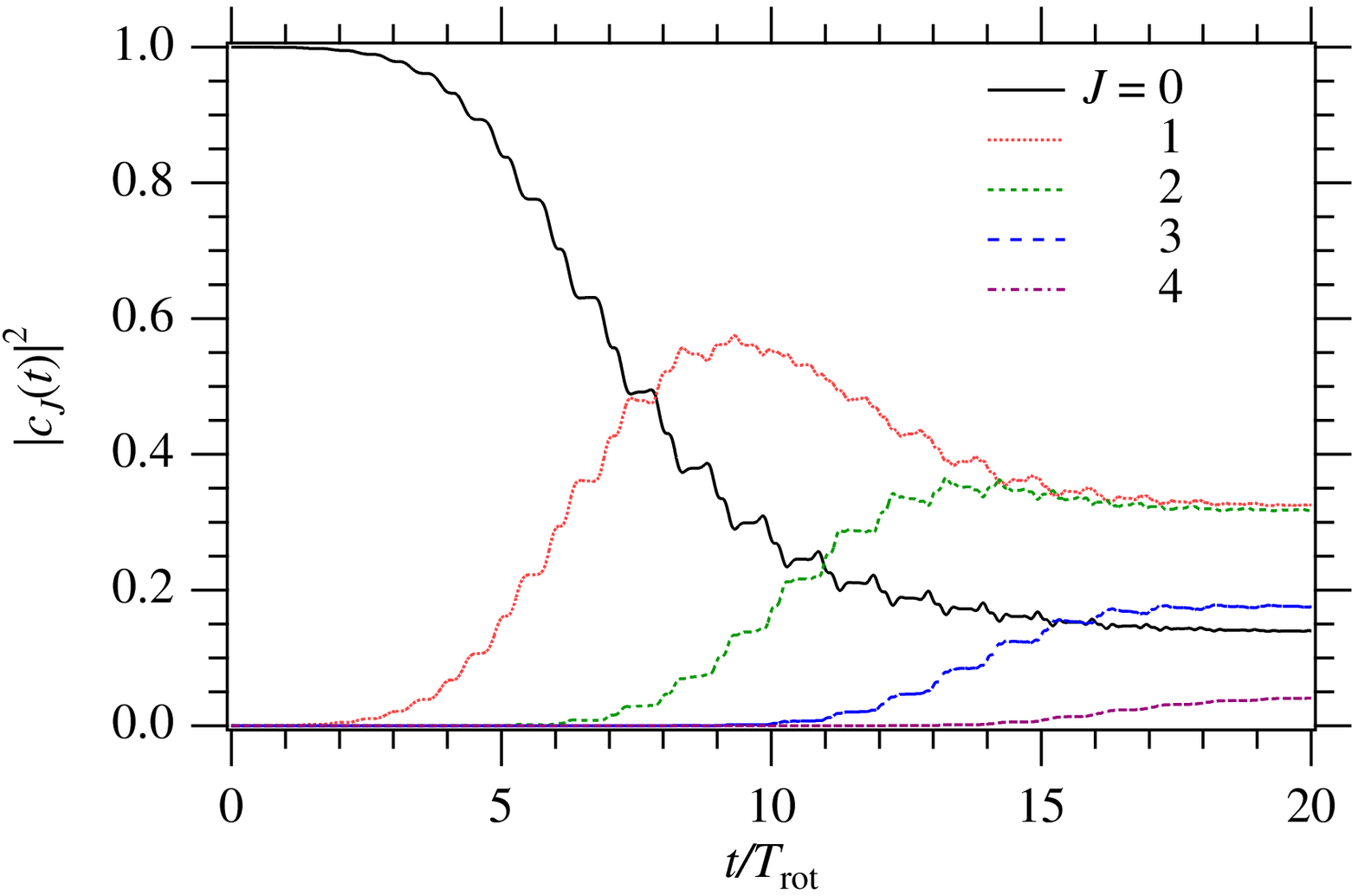}
\caption{Time evolution of the population of rotational states of a
  rigid rotor interacting with the electric field given in
  Fig.~\ref{fig:champ_cos}(a).}
\label{fig:coeff_cos}
\end{figure}

\begin{figure}
\includegraphics[width=\figwidth]{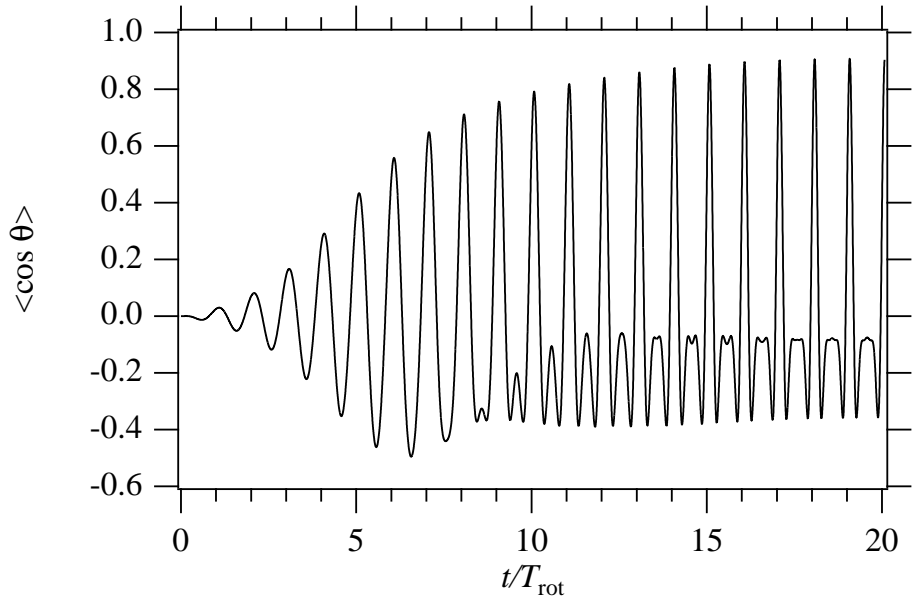}
\caption{Orientation, as measured by $\cth$, obtained for a rigid
  rotor interacting with the electric field given in
  Fig.~\ref{fig:champ_cos}(a).  The field-free evolution is then
  periodic with period $T_{\mathrm{rot}}$.}
\label{fig:cosavg_cos}
\end{figure}

The electric field obtained using the cost functional $\mathcal{J}_1$
[Eq.~(\ref{eq:costop})] with the observable $\hat{O} = \cos\theta +
\hat{I}$, i.e., for the optimization of the orientation, is given in
Fig.~\ref{fig:champ_cos}(a). To better analyze the result, we have
performed a short-time Fourier transform
(STFT),\cite{math:priestley81}
 \begin{equation}\label{sfft}
\mathcal{F}(\nu,t)=\int_{-\infty}^{+\infty}
\mathcal{E}(\tau) w(\tau-t)  e^{-i 2\pi \nu\tau} d\tau,
\end{equation}
where $w$ is a Tukey-Hanning window with a temporal width of $1.9
\times 10^6\ \mathrm{a.u.}$ The frequency distribution $\mathcal{F}$
can be seen in Fig.~\ref{fig:champ_cos}(b), where the ordinate is the
dimensionless value $2 \nu T_{\mathrm{rot}}$, corresponding to the
dimensionless energy $E/B$.  The energy spacing between rotational
states $J$ and $J+1$ being $2 B J$, we see clearly from
Fig.~\ref{fig:champ_cos}(b) that the field is initially resonant with
the $J = 0 \rightarrow 1$ transition, and subsequently comes in
resonance with higher and higher pairs of rotational levels $J = 1
\rightarrow 2$, $J = 2 \rightarrow 3$, \dots\@ Looking at the
population of the rotational states, in Fig.~\ref{fig:coeff_cos}, we
indeed find that, starting from the ground state $J = 0$, the molecule
is pumped to the first excited state, then to the second, etc.  At the
end of interaction with the field, the population distribution is such
that an orientation of $\cth(t_\mathrm{f}) = 0.909$ is attained
(Fig.~\ref{fig:cosavg_cos}).  In other words, the molecule is oriented
by a process of \emph{rotational ladder climbing}.

\begin{figure}
\includegraphics[width=\figwidth]{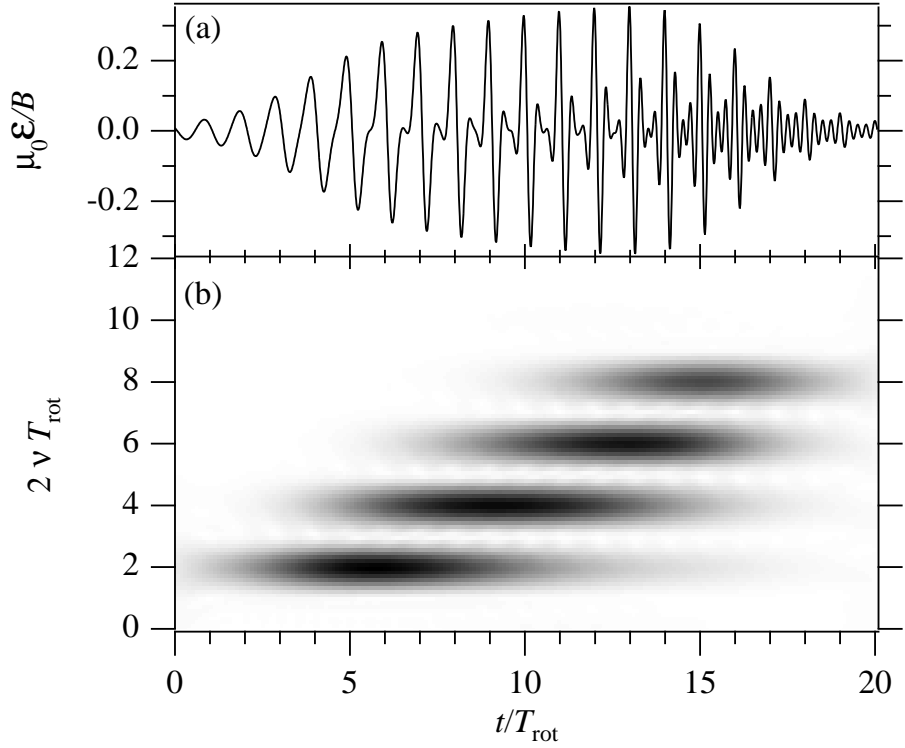}
\caption{(a) Electric field obtained with criterion $\mathcal{J}_2$
  for the optimization of the 
  projection of wave function on the target $\cible$ corresponding to
  orientation (see Table~\ref{tab:target}).
  (b) Short-time Fourier transform of the field in (a).}
\label{fig:champ_cible1}
\end{figure}

\begin{figure}
\includegraphics[width=\figwidth]{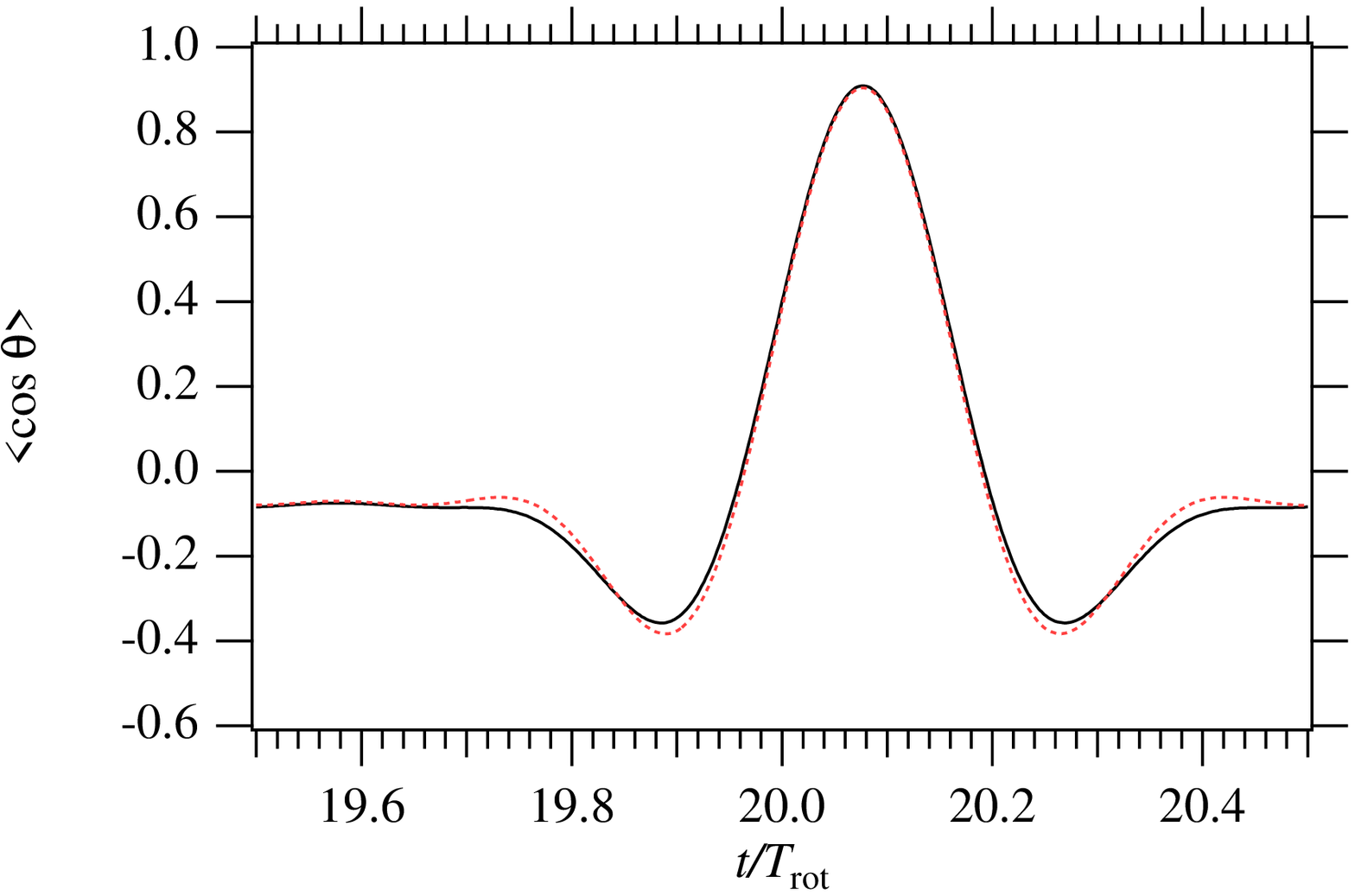}
\caption{Orientation, as measured by $\cth$, obtained for the
  interaction with the electric field given in
  Fig.~\ref{fig:champ_cos}(a) (solid line) and
  Fig.~\ref{fig:champ_cible1}(a) (dashed line).  The field-free
  evolution is then periodic with period $T_{\mathrm{rot}}$.}
\label{fig:cosavg}
\end{figure}

If we instead take the cost functional $\mathcal{J}_2$
[Eq.~(\ref{eq:costop_target})] with $\cible$ the target state for
orientation given in Table~\ref{tab:target}, we obtain a result very
similar to the previous one, as shown in Fig.~\ref{fig:champ_cible1}.
The main difference is the absence of the frequency component at $2
\nu T_{\mathrm{rot}} = 10$, which is easily understood from the fact
that it corresponds to the $J = 4 \rightarrow 5$ transition, while the
target is restricted to $J_{\mathrm{max}} = 4$.  The resulting
dynamics of $\cth$ are nearly indistinguishable, as is seen in
Fig.~\ref{fig:cosavg}. The similarity of both results can also be
explained by looking at the projection on the target $P \equiv \left|
  \left\langle \cible \left| \psi(t_\mathrm{f}) \right\rangle
  \right. \right|^2$.  For the wave function obtained for the
optimization of the observable, we already have $P=0.9933$, the
optimization of the projection on the target allowing to reach $P =
0.9969$.  This efficiency is better than that obtained when kicking
the molecule with short pulses,\cite{align:sugny04a,orient:dion05a}
but the time necessary to reach the optimized state is of the order of
20 rotational periods, while it takes less than one molecular rotation
with kicks.  The time needed for ladder climbing also explains why the
state reached for the optimization of $\cth$ is almost the same as the
target state, since the time limit imposed by $t_\mathrm{f}$
constrains the maximum value of $J$ than can be excited.  It can be
likened to the reduced Hilbert space used when defining the target
state.\cite{align:sugny04a,control:sugny05a}

\subsection{Optimizing alignment}

\begin{figure}
\includegraphics[width=\figwidth]{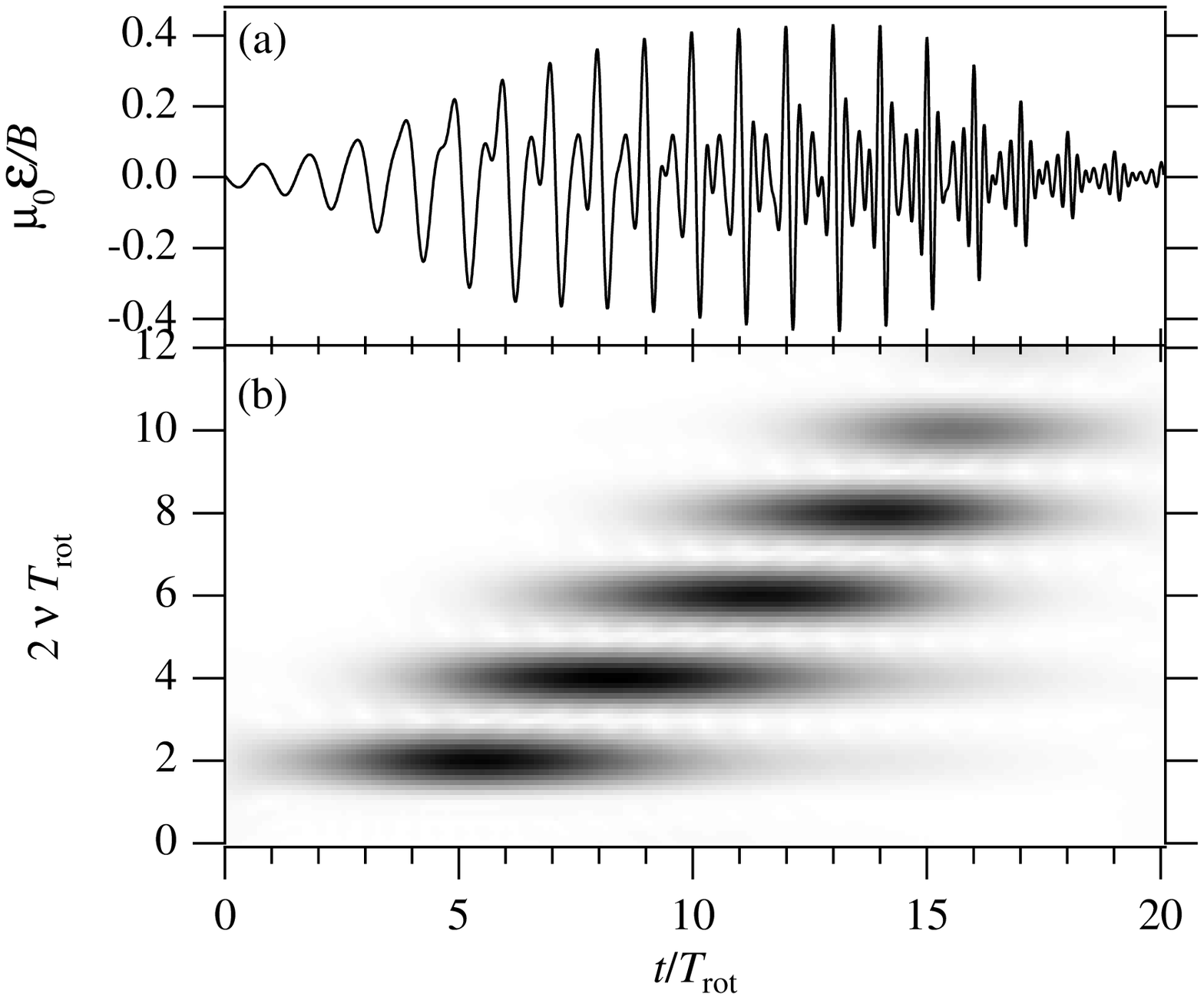}
\caption{Same as Fig.~\ref{fig:champ_cos}, but for the optimization of
  $\csth$.} 
\label{fig:champ_cos2}
\end{figure}

\begin{figure}
\includegraphics[width=\figwidth]{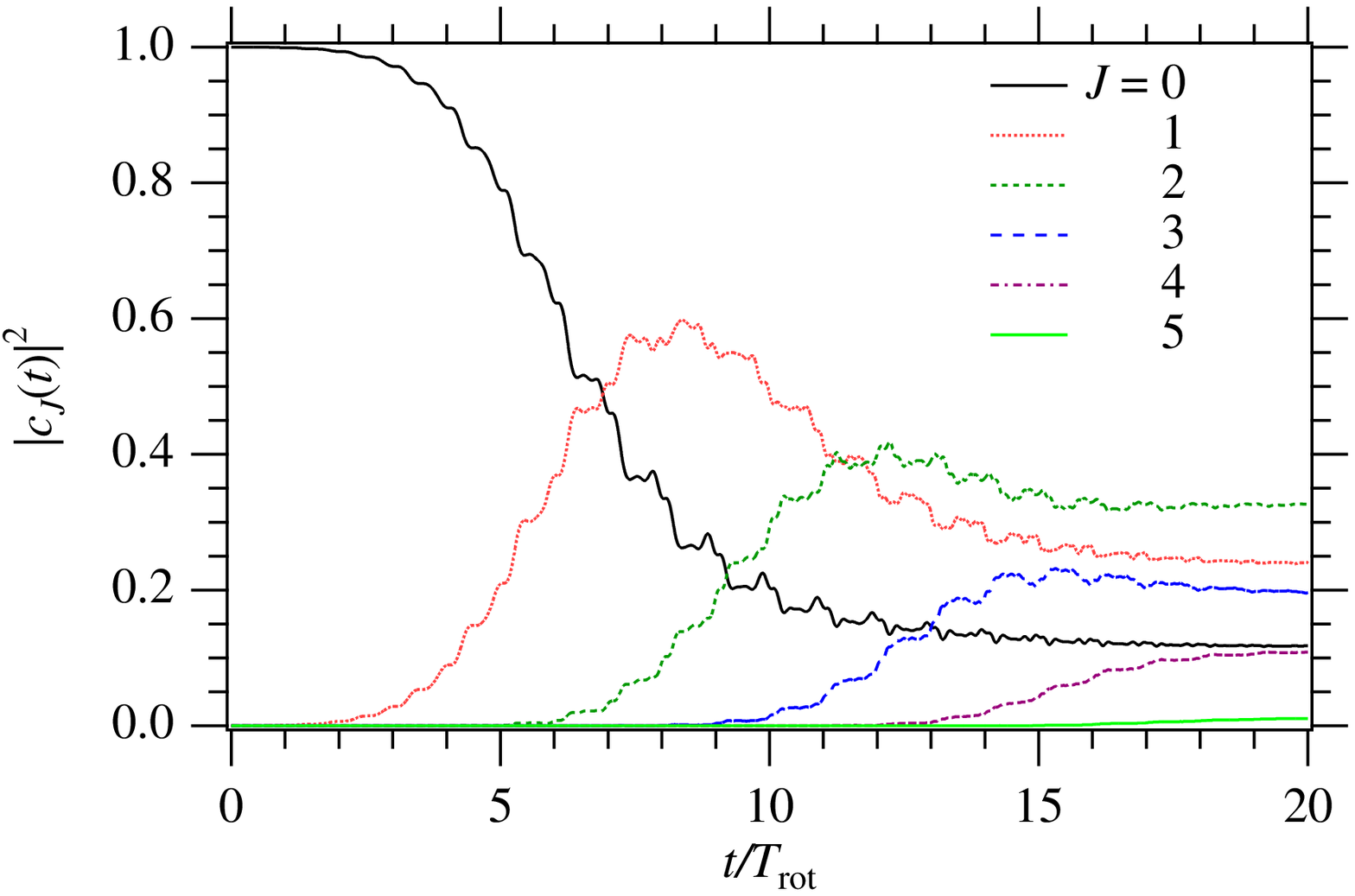}
\caption{Time evolution of the population of rotational states of a
  rigid rotor interacting with the electric field given in
  Fig.~\ref{fig:champ_cos2}(a).}
\label{fig:coeff_cos2}
\end{figure}

The result for the maximization of the operator $\hat{O} = \cos^2
\theta + \hat{I}$ to achieve alignment is given in
Fig.~\ref{fig:champ_cos2}.  The field obtained is almost the same as
the one obtained for orientation, except that the rotational
excitation happens at a quicker pace, as displayed in
Fig.~\ref{fig:coeff_cos2}, where it seen that $J = 5$ is now
significantly populated.  The alignment obtained is $\csth = 0.866$.

\begin{figure}
\includegraphics[width=\figwidth]{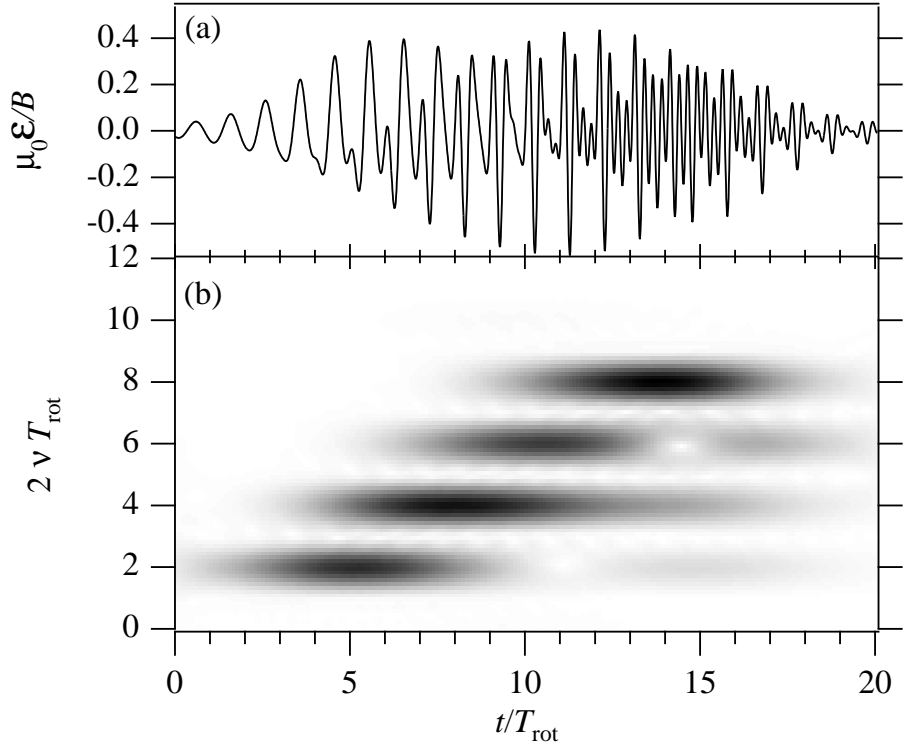}
\caption{Same as Fig.~\ref{fig:champ_cible1}, but for the optimization
  of the projection of wave function on the target $\cible$
  corresponding to alignment (see Table~\ref{tab:target}).}
\label{fig:champ_cible2}
\end{figure}

\begin{figure}
\includegraphics[width=\figwidth]{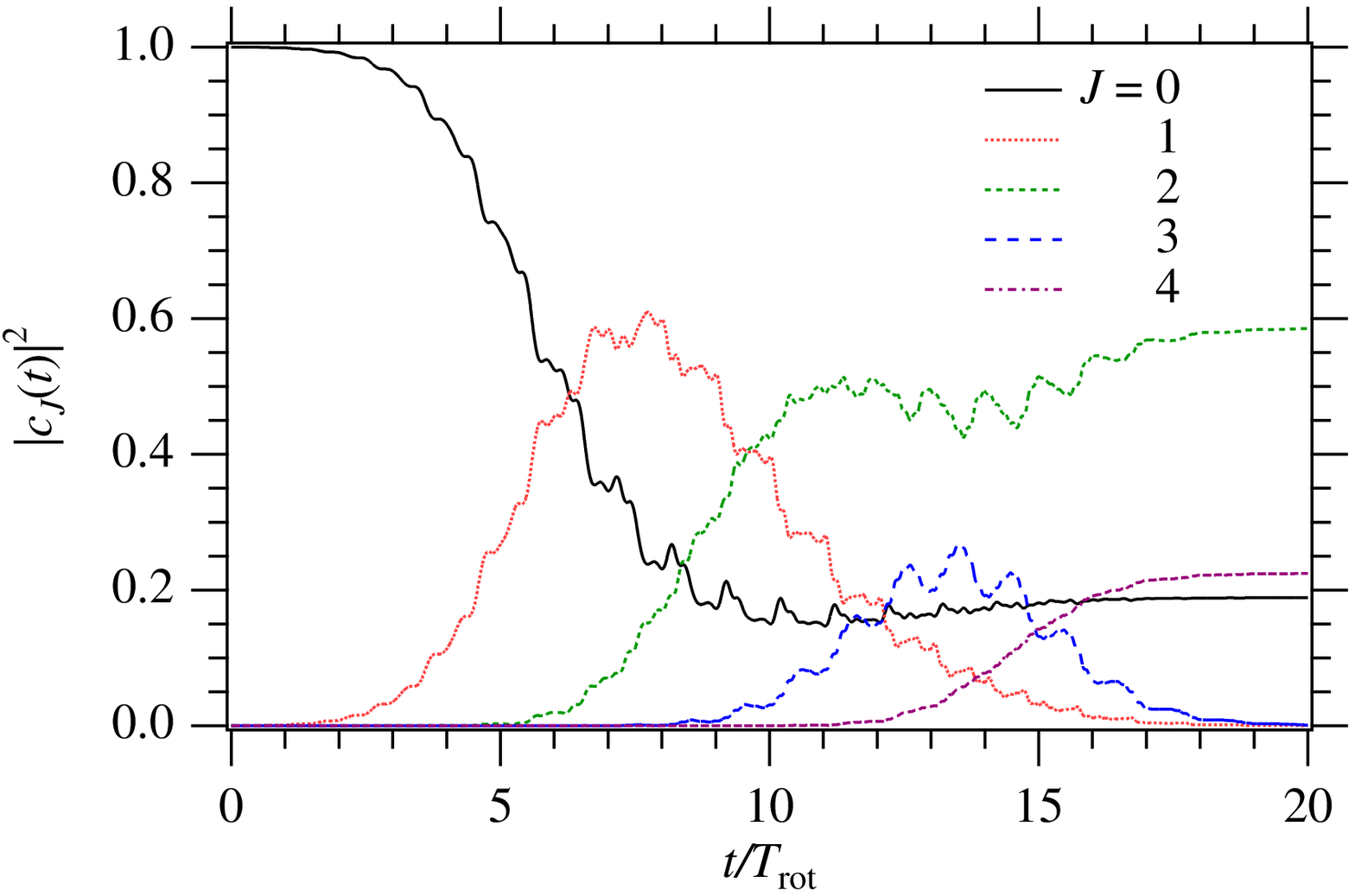}
\caption{Time evolution of the population of rotational states of a
  rigid rotor interacting with the electric field given in
  Fig.~\ref{fig:champ_cible2}(a).}
\label{fig:coeff_cible2}
\end{figure}

Changing now the the target state for alignment
(Table~\ref{tab:target}), we see in Fig.~\ref{fig:champ_cible2} that
the field obtained is significantly different.  The frequency
component corresponding to an energy of $E = 4B$ is present for a
longer time and components at $E = 6B$ and, to a lesser extent, $E =
2B$, reappear near the turn-off of the field. The time
dependence of the population of the rotational states in
Fig.~\ref{fig:coeff_cible2} gives the explanation of this phenomenon:
the populations of $J=1$ and 3 are \emph{pumped down} by these later
components, since only even $J$ levels are populated in the optimally
aligned target state.  The original excitation to the odd levels was
necessary as the rotational states are only significantly coupled via
the permanent dipole moment, implying the selection rule $\Delta J =
\pm 1$, the role of the polarizability being here negligible.  This
excitation-deexcitation scheme leads to a projection on the target
state of $P=0.9950$, compared with $P=0.5487$ when only optimizing for
alignment. 

\begin{figure}
\includegraphics[width=\figwidth]{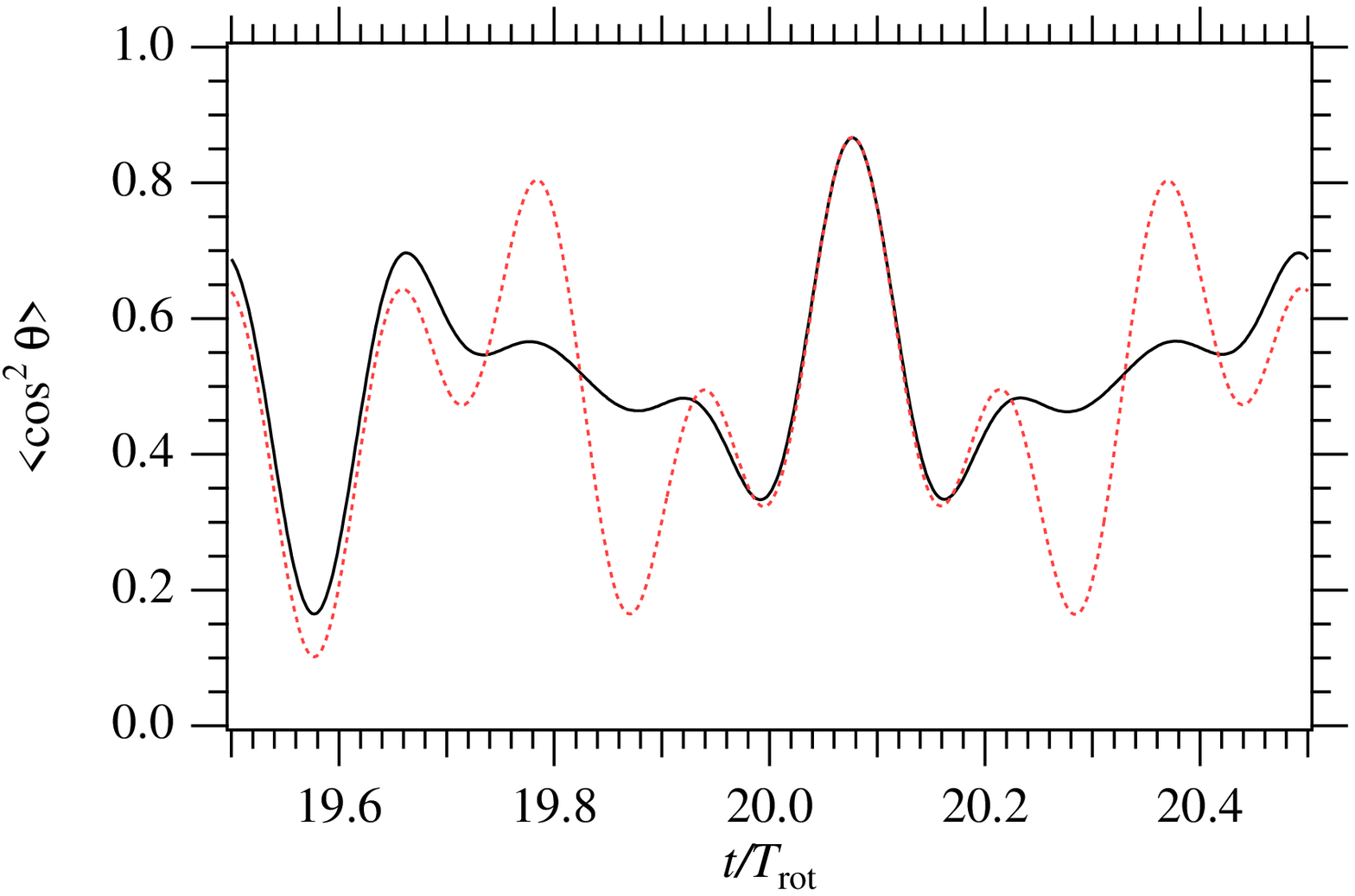}
\caption{Alignment, as measured by $\csth$, obtained for the
  interaction with the electric field given in
  Fig.~\ref{fig:champ_cos2}(a) (solid line) and
  Fig.~\ref{fig:champ_cible2}(a) (dashed line).  The field-free
  evolution is then periodic with period $T_{\mathrm{rot}}$.}
\label{fig:cos2avg}
\end{figure}

It is interesting to note that the maximum alignment obtained is the
same as in the first case, with $\csth(t_\mathrm{f}) = 0.867$
(Fig.~\ref{fig:cos2avg}), even though the two wave functions obtained
involve very different mixtures of spherical harmonics.  One striking
contrast between the two is actually not visible when looking only at
$\csth$: in the second case, the state obtained is \emph{strictly}
aligned, in the sense that the angular distribution is symmetric with
respect to $\theta = \pi/2$.  In the first case, the maximum in
alignment corresponds also to a maximum in orientation, with
$\cth(t_\mathrm{f}) = 0.891$, whereas in the second case
$\cth(t_\mathrm{f}) = -0.027$.

\subsection{Considering a rotational temperature}

\begin{figure}
\includegraphics[width=\figwidth]{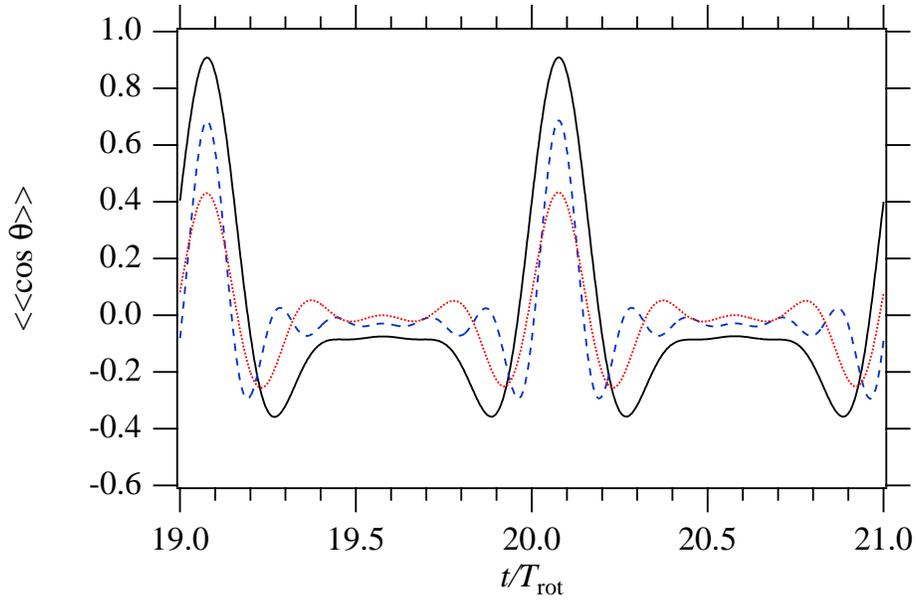}
\caption{Orientation, as measured by $\left\langle \cth
  \right\rangle$, obtained for a rigid rotor interacting with the
  electric field given in Fig.~\ref{fig:champ_cos}(a), starting from
  the rotational ground state (solid line) and from a rotational
  temperature of $k_\mathrm{B} T/B \approx 4.77$ (dotted line).  For
  comparison, the orientation obtained with the field optimized for
  this temperature [Fig.~\ref{fig:champ_cos_temp}(a)] is also shown
  (dashed line).}
\label{fig:cosavg_temp}
\end{figure}

By starting all simulations from the ground rotational state, we have
in fact made the approximation of a zero initial rotational
temperature. From previous work on laser-induced alignment and
orientation,\cite{orient:dion02,orient:machholm01,align:ortigoso99,%
  align:seideman01,align:machholm01} it is known that considering a
higher, experimentally more realistic initial rotational temperature
will lead to an important decrease in the amount of
orientation/alignment obtained.  This can be seen in
Fig.~\ref{fig:cosavg_temp}, where we show how orientation is affected
when the optimal field presented in Fig.~\ref{fig:champ_cos}(a) is
applied to a thermal ensemble of initial temperature $k_\mathrm{B} T/B
\approx 4.77$ (corresponding to 10~K for HCN).

\begin{figure}
\includegraphics[width=\figwidth]{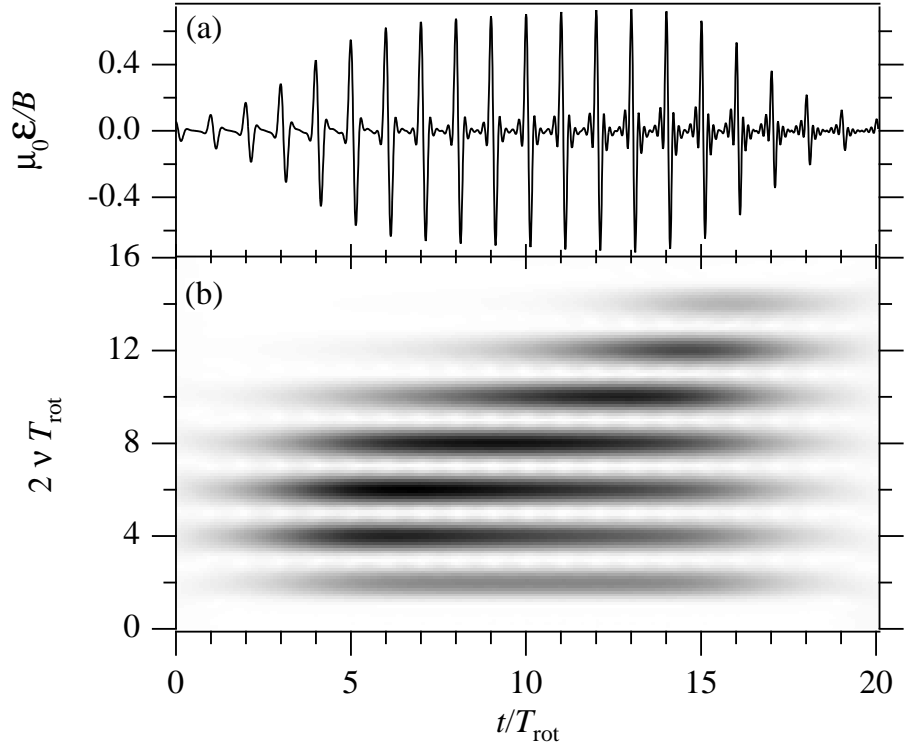}
\caption{(a) Electric field obtained for the optimization of
  $\left\langle \cth \right\rangle$ starting from a rotational
  temperature of $k_\mathrm{B} T/B \approx 4.77$. (b) Short-time
  Fourier transform of the field in (a).}
\label{fig:champ_cos_temp}
\end{figure}

This difficulty can be overcome by adapting the optimization method to
take into account the initial thermal
distribution.\cite{orient:benhajyedder02,control:turinici04} We
present here the optimization of orientation, which can be trivially
extended to the case of alignment or of a suitable mixed-state
target.\cite{control:sugny05b} The measure of orientation now reads
\begin{equation}
\left\langle \cth \right\rangle(t) = Q^{-1} \sum_{J=0}^{\infty} \exp
\left[ \frac{-B J(J+1)}{k_\mathrm{B} T} \right] \sum_{M=-J}^{J}
\cth_{J,M}(t), 
\label{eq:cthT}
\end{equation}
where 
\begin{equation}
Q = \sum_{J=0}^{\infty} (2J + 1) \exp \left[ \frac{-B
    J(J+1)}{k_\mathrm{B} T} \right] 
\end{equation}
is the partition function and $\cth_{J,M}(t) \equiv \left\langle
  \psi_{J,M}(\theta;t) \right| \cos\theta \left| \psi_{J,M}(\theta;t)
\right\rangle $ is obtained by solving the TDSE with the
Hamiltonian~(\ref{eq:tdse}) for the initial condition
$\psi_{J,M}(\theta;t=0) \equiv Y_{J,M}(\theta)$.  Considering a
temperature $k_\mathrm{B} T/B \approx 4.77$, we can restrict the sum
over $J$ in Eq.~(\ref{eq:cthT}) to $J_\mathrm{max} = 7$, with 16 basis
functions [see Eq.~(\ref{eq:expand})] used for the time evolution.
The monotonic algorithm~(\ref{eq:algo}) becomes
\begin{equation}
\left\{
\begin{array}{rcl}
  \mathcal{E}^{k+1}(t) & = & - Q^{-1} {\displaystyle
 \sum_{J=0}^{J_\mathrm{max}}} \exp
\left[ \frac{-B J(J+1)}{k_\mathrm{B} T} \right] {\displaystyle \sum_{M=-J}^{J}}
\frac{\Im \left\langle \psi_{J,M}^{k+1}(t) \right|
    \mu_0\cos\theta \left| \chi_{J,M}^k(t) \right\rangle}{2\Im \left\langle
      \psi_{J,M}^{k+1}(t) \right| \frac{\Delta\alpha\cos^2\theta
      +\alpha_{\perp}}{2} \left| \chi_{J,M}^k(t) \right\rangle -\lambda(t)},
  \\
  i\partial_t \psi_{J,M}^{k+1}(t) & = & \left[
  B-\mu_0\mathcal{E}^{k+1}(t)\cos\theta
  -\frac{[\mathcal{E}^{k+1}(t)]^2}{2}
  \left(\Delta\alpha\cos^2\theta+\alpha_{\perp} \right) \right]
  \psi_{J,M}^{k+1}(t), 
  \\
  \psi_{J,M}^{k+1}(\theta,t=0) & = & Y_{J,M}(\theta).
\end{array}\right.
\end{equation}
The resulting optimized field, given in
Fig.~\ref{fig:champ_cos_temp}(a), allows to reach $\left\langle \cth
\right\rangle(t_\mathrm{f}) = 0.686$, which is much better than the
value of $0.434$ obtained when the $T=0\ \mathrm{K}$ field is applied
to the same thermal distribution, as shown in
Fig.~\ref{fig:cosavg_temp}.  Looking at the Fourier transform of the
field, Fig.~\ref{fig:champ_cos_temp}(b), we see that the mechanism to
achieve orientation is slightly different, with the lower frequencies
present for a much longer time than previously observed, which
reflects the fact that many $J\neq 0$ states are initially populated.
Nevertheless, the ladder climbing structure of the resonances is still
present.

\section{Conclusion}

Using a monotonically convergent algorithm, we have searched for the
optimal electric field maximizing either the alignment or the
orientation of a linear molecule, taken in a rigid rotor
approximation.  We have carried out the optimization both in terms of
maximization of observables corresponding to orientation/alignment and
using target states offering a good compromise between the efficiency
of orientation/alignment and its duration.

We have found that, starting from the ground rotational state, the
optimal fields allow to reach orientation/alignment by rotational
ladder climbing, i.e., by successive resonant excitation of
neighboring rotational levels.  We insist on the fact that this
scenario appears ``naturally'' from the physics of the problem and is
not imposed \emph{a priori} by the optimization algorithm.  This
process allows to reach an orientation of $\cth = 0.909$ or an
alignment of $\csth = 0.867$.  Target states can also be reached to
within better than 0.5\%.

We have also shown how our optimization method can be extended to the
more realistic case of an initial thermal distribution of rotational
states.  Using orientation as an illustrative example, we obtained a
value of $\left\langle \cth \right\rangle = 0.686$ at a temperature
$k_\mathrm{B} T/B \approx 4.77$ (corresponding to 10~K for HCN).  This
result is much better than those previously
reported,\cite{orient:benhajyedder02} even though we are considering
here a \emph{higher} rotational temperature.

As a rigid rotor model was used, this study did not take into account
any vibrational excitation, which could hinder or enhance the
orientation/alignment obtained. By including vibrations into the
model, it is possible to use different control paths not involving
direct rotational excitation, enabling the choice of infra-red lasers
as control fields.\cite{align:dion99,orient:dion99,align:hoki01}
Vibration-rotation coupling can also lead to cross-revivals of
vibrational wave packets.\cite{rot:hansson00,rot:wallentowitz02}
Future work will thus take into account the vibration of the molecule.

\begin{acknowledgments}
  We thank Yvon Maday and Arne Keller for stimulating discussions.
  Financial support from the \emph{Action Concert\'{e}e Incitative
    Nouvelles Interfaces des Math\'{e}matiques} is gratefully
  acknowledged.
\end{acknowledgments}

\clearpage


\begin{table}
\begin{center}
\begin{ruledtabular}
\begin{tabular}{|c|l|l|} 
$J$ & \multicolumn{1}{c|}{$c_{J}^{\mathrm{o}}$} &
\multicolumn{1}{c|}{$c_{J}^{\mathrm{a}}$} \\ \hline 
0 & 0.344185 & 0.413914 \\
1 & 0.540216 & 0. \\
2 & 0.563165 & 0.744364 \\
3 & 0.456253 & 0. \\       
4 & 0.253736 & 0.524021 
\end{tabular}
\end{ruledtabular}
\end{center}
\caption{Expansion coefficients [see Eq.~(\ref{eq:expand})] for the
  target states $\cible$ corresponding to maximum orientation,
  $c_{J}^{\mathrm{o}}$, and alignment, $c_{J}^{\mathrm{a}}$, when the
  rotational excitation is restricted to $J_{\mathrm{max}} = 4$.}
\label{tab:target}
\end{table}


\end{document}